\documentclass[pre,groupedaddress,superscriptaddress,reprint,twocolumn,showpacs,notitlepage,nofootinbib]{revtex4}

\usepackage{amsfonts}
\usepackage{amsmath}
\usepackage{amssymb}
\usepackage{graphicx}
\usepackage{hyperref}
\usepackage{color}
\usepackage{mathrsfs}
\usepackage{bbm}
\usepackage{subfigure}

\usepackage{times,txfonts}
\usepackage{calrsfs}

\usepackage{siunitx}

\newcommand{\ignore}[1]{}
\newcommand{\nobibentry}[1]{{\let\nocite\ignore\bibentry{#1}}}
\newcommand{\bibfnamefont}[1]{#1}
\newcommand{\bibnamefont}[1]{#1}

\newcommand{\ket}[1]{\left\vert#1\right\rangle}
\newcommand{\bra}[1]{\left\langle#1\right\vert}

\begin{document}

\title{Optimal Performance of Endoreversible Quantum Refrigerators}

\author{Luis A. Correa}
\email{luis.correa@nottingham.ac.uk}
\affiliation{School of Mathematical Sciences, The University of Nottingham, University Park, Nottingham NG7 2RD, UK}
\affiliation{IUdEA Instituto Universitario de Estudios Avanzados, Universidad de La Laguna, 38203 Spain}

\author{Jos\'{e} P. Palao}
\affiliation{IUdEA Instituto Universitario de Estudios Avanzados, Universidad de La Laguna, 38203 Spain}
\affiliation{Departamento de F\'{i}sica, Universidad de La Laguna, 38204 Spain}

\author{Gerardo Adesso}
\affiliation{School of Mathematical Sciences, The University of Nottingham, University Park, Nottingham NG7 2RD, UK}

\author{Daniel Alonso}
\affiliation{IUdEA Instituto Universitario de Estudios Avanzados, Universidad de La Laguna, 38203 Spain}
\affiliation{Departamento de F\'{i}sica, Universidad de La Laguna, 38204 Spain}

\pacs{05.70.Ln, 03.65.-w, 05.40.-a}
\date{\today}

\begin{abstract}
The derivation of general performance benchmarks is important in the design of highly optimized heat engines and refrigerators. To obtain them, one may model phenomenologically the leading sources of irreversibility ending up with results which are model-independent, but limited in scope.  Alternatively, one can take a simple physical system realizing a thermodynamic cycle and assess its optimal operation from a complete microscopic description. We follow this approach in order to derive the coefficient of performance at maximum cooling rate for \textit{any} endoreversible quantum refrigerator. At striking variance with the \textit{universality} of the optimal efficiency of heat engines, we find that the cooling performance at maximum power is crucially determined by the details of the specific system-bath interaction mechanism. A closed analytical benchmark is found for endoreversible refrigerators weakly coupled to unstructured bosonic heat baths: an ubiquitous case study in quantum thermodynamics.
\end{abstract}

\maketitle

\section{Introduction}\label{sec1}

Energy conversion systems, including heat engines and refrigerators, encompass a broad variety of devices which find widespread uses in the domestic, industrial and academic domains. Design optimization of such systems is crucial for their implementation to be cost-efficient, and the determination of general performance benchmarks to assess their `optimality', is a very active research area \cite{andresen2011current,hoffmann1997endoreversible}. 
A familiar example of heat engine is a nuclear power station. 
The relevant figure of merit to benchmark its optimality is the output power rather than the efficiency of energy conversion \cite{novikov1957efficiency}: 
In fact, capital costs are by far the dominant contribution to the price of the kWh, while the nuclear fuel itself is comparatively inexpensive. Hence, ideally, a nuclear energy station will be designed to operate at the maximum power output $\mathcal{P}_*$ corresponding to some heat input $\dot{\mathcal{Q}}_{h,*}$, which defines an \textit{optimal efficiency} $\eta_*\equiv-\mathcal{P}_*/\dot{\mathcal{Q}}_{h\,*}$. 

As a working assumption, one may treat a nuclear power station as a perfect Carnot engine running between heat reservoirs at temperatures $T_c<T_h'~(<T_h)$, where $T_h'$ is the effective temperature of the working fluid at the hot end of the cycle. This amounts to saying that the leading source of irreversibility in atomic power generation is the imperfect thermal contact of the working fluid with the reactor, to the point that internal friction and heat leaks may be completely disregarded. This is known as \textit{endoreversible} approximation \cite{hoffmann1997endoreversible}. If one further assumes a simple Newtonian heat transfer law for the heat current $\dot{\mathcal{Q}}_h=C_h(T_h-T_h')$, where $C_h$ is a constant, then the effective temperature maximizing the power may be found to be the geometric mean of $T_h$ and $T_c$. Consequently, the optimal efficiency reads
\begin{equation}
\eta_*=1-T_c/T_{h\,*}'=1-\sqrt{T_c/T_h}=1-\sqrt{1-\eta_C},
\label{curzon_ahlborn}
\end{equation}
where $\eta_C=1-T_c/T_h$ is the ultimate Carnot efficiency \cite{carnot1890fire}. This formula, introduced by Yvon \cite{Yvon1955reactor} and Novikov \cite{novikov1957efficiency} in the mid 1950s in the context of atomic energy generation, was re-derived twenty years later by Curzon and Ahlborn \cite{curzon1975efficiency} in their 1975 seminal paper \footnote{Interestingly, the origins of Eq.~\eqref{curzon_ahlborn} can be traced back to a book by H. B. Reitlinger, first published in 1929 \cite{1406.5853v2}.}. In principle, it should be nothing but a crude approximation to optimality, but it turns out to be in good agreement with the observed efficiency of actual thermal power plants, and proves to be remarkably independent of the specific design \cite{Yvon1955reactor}. Indeed, it agrees with the optimal efficiency of any engine operating close to equilibrium \cite{PhysRevLett.95.190602,esposito2009universality}, and applies quite generally to symmetric low-dissipation engines \cite{PhysRevLett.105.150603}, even if these are realized on a quantum mechanical support \cite{esposito2010universalityCA,geva1991spin}. Eq.~\eqref{curzon_ahlborn} is, therefore, a useful design guideline, as it reliably benchmarks the optimal operation of a large class of heat engines. Besides, it is clearly model-independent.

In the last few decades, many attempts have been made to answer the fundamental question of whether a similar model-independent benchmark can be obtained for optimal cooling. That would certainly be very useful in the design optimization of refrigerators, but unfortunately the straightforward endoreversible approach together with the assumption of a linear heat transfer law does not help in this case: The cooling rate $\dot{\mathcal{Q}}_c$, which replaces $\mathcal{P}$ as figure of merit, is maximal only at vanishing \textit{coefficient of performance} (COP) $\varepsilon\equiv\dot{\mathcal{Q}}_c/\dot{\mathcal{P}}$. This problem might be circumvented by resorting to alternative heat transfer laws, though these usually lead to involved (non-universal) formulas for the optimal COP, explicitly depending on phenomenological heat conductivities \cite{0022-3727_23_2_002}. 

Benchmarks analogous to Eq.~\eqref{curzon_ahlborn}, may still be obtained by retaining the simple Newtonian ansatz and changing instead the definition of `optimality'. Practical considerations may advise e.g.~to pay the same attention to the COP and the cooling rate, so that the meaningful figure of merit would be $\chi\equiv\varepsilon~\dot{\mathcal{Q}}_c$ rather than $\dot{\mathcal{Q}}_c$ alone. In this case, one would find an optimal performance of $\varepsilon_*=\sqrt{1+\varepsilon_C}-1$ \cite{0022-3727_23_2_002}, which holds in fact for any symmetric low-dissipation Carnot refrigerator \cite{PhysRevE.85.010104}. Here, $\varepsilon_C=T_c/(T_h-T_c)$ stands for the Carnot COP. Other criteria for optimality \cite{PhysRevLett.78.3241,PhysRevE.73.057103} would lead, of course, to different performance benchmarks\footnote{Another option would be to relax the endoreversible approximation, allowing for heat leaks and internal friction, while keeping $\dot{\mathcal{Q}}_c$ as figure of merit, and a simple linear model for the heat currents \cite{chen1994new,0295-5075-103-4-40001}. Generally, this also leads to model-dependent benchmarks.}. 

In this paper, we analyze the COP at maximum cooling rate for endoreversible quantum refrigerators, generally modelled as \textit{tricycles} \cite{1310.0683v1}. We find that the details of the system-bath interaction mechanism place a tight upper bound on the cooling performance, which automatically precludes the derivation of any model-independent benchmarks. We then look into the paradigmatic case of a three-level compression refrigerator \cite{PhysRevLett.2.262,PhysRevE.64.056130} operating between unstructured bosonic heat baths, to obtain a simple closed expression for $\varepsilon_*(\varepsilon_C)$, which is further shown to bound and closely reproduce the optimal COP of \textit{any} multi-stage endoreversible refrigerator within the same dissipative scheme. Our analysis unveils fundamental differences between heat engines and refrigerators from the point of view of their optimal performance, and highlights the key importance of reservoir engineering in the optimization of technologically relevant quantum models.

This paper is structured as follows: The generic template of a quantum tricycle is briefly described Sec.~\ref{sec2}. Then, our main result, concerning the non-universality of the optimal cooling performance is derived in Sec.~\ref{sec3} and illustrated with a simple example in Sec. \ref{sec4}. Finally, in Sec.~\ref{sec5} we summarize and draw our conclusions. For the sake of clarity, the technical details of the derivation of quantum master equations for periodically-driven systems are postponed until Appendix \ref{appendix}.

\begin{figure}
	\includegraphics[width=5.5cm]{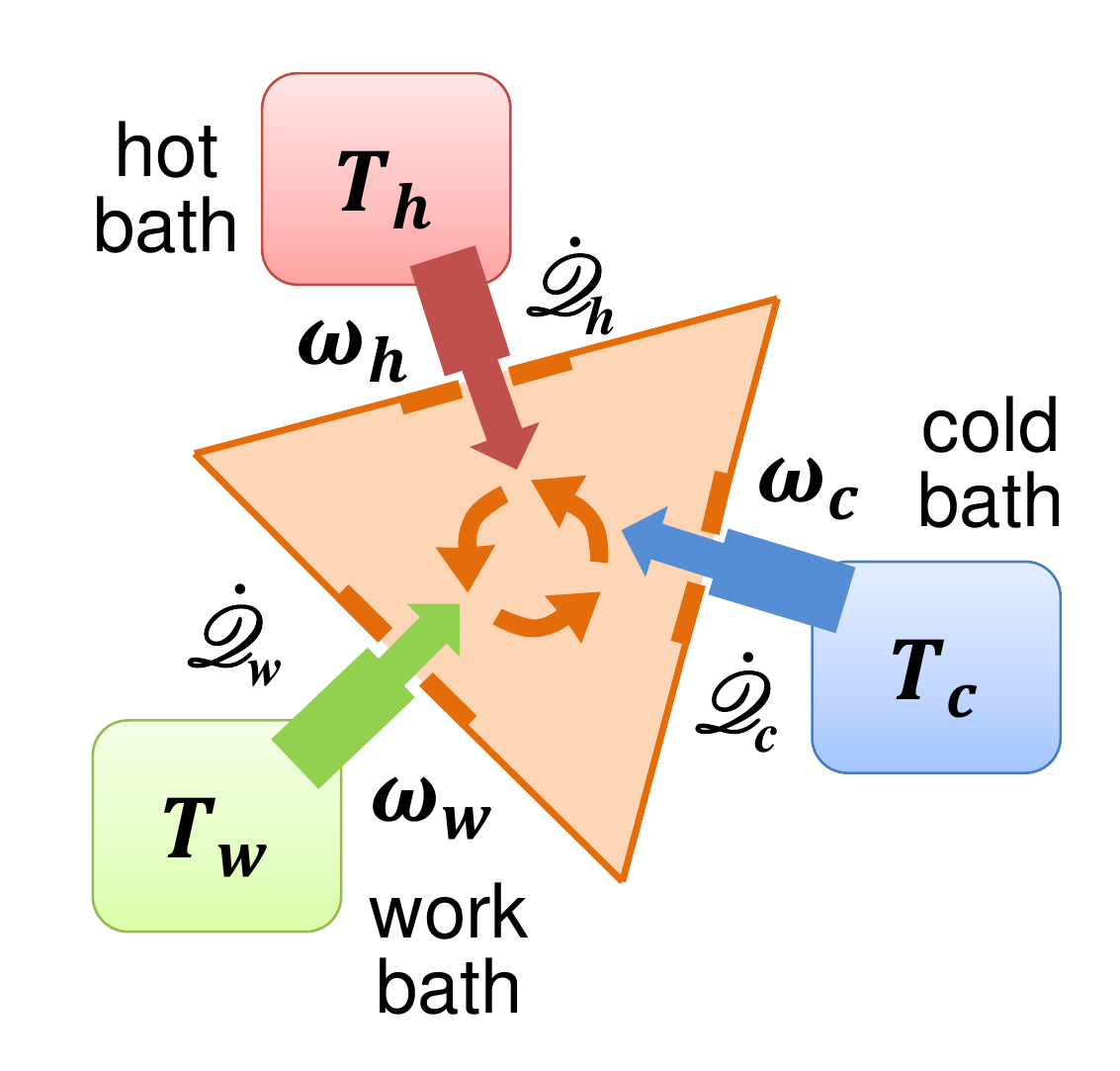}
\caption{Quantum tricycle: A quantum system selectively coupled through frequency filters to three heat baths (with temperatures $T_w > T_h > T_c$), embodies the prototype of any thermal device. Here, the direction of the heat currents (depicted as arrows) correspond to a refrigerator. Reversing them realizes a heat transformer/heat engine.}
\label{fig1}
\end{figure}

\section{Endoreversible quantum tricycles}\label{sec2} 

A generic energy conversion device may be thought of as a stationary black box in simultaneous thermal contact with three heat reservoirs at different temperatures $T_w>T_h>T_c$ or, alternatively, with two heat reservoirs $T_h>T_c$ and a work repository ($T_w\rightarrow\infty$), which, in principle, accounts for the case of a heat engine or a power-driven refrigerator (we shall elaborate more on this equivalence in an example below). This template, termed `tricycle' \cite{andresen1976tricycle}, is suitable to describe averaged finite-time cycles or continuous processes, and is represented by the triple $\{\dot{\mathcal{Q}}_w,\dot{\mathcal{Q}}_h,\dot{\mathcal{Q}}_c\}$ of steady-state rates of incoming (positive) and outgoing (negative) energy flow in the system through each of the thermal contact ports. In order to comply with the first and second laws of thermodynamics, these must satisfy
\begin{subequations}\label{thermo_laws}
\begin{align}
\dot{\mathcal{Q}}_w+\dot{\mathcal{Q}}_h+\dot{\mathcal{Q}}_c&=0 \label{first_law} \\
\frac{\dot{\mathcal{Q}}_w}{T_w}+\frac{\dot{\mathcal{Q}}_h}{T_h}+\frac{\dot{\mathcal{Q}}_c}{T_c}&\equiv-\dot{S}\leq 0. \label{second_law}
\end{align}
\end{subequations}

If the black box encloses a quantum system, thermal contact with the heat reservoir may be selectively established through filters at frequencies $\omega_\alpha$ with $\alpha\in\{w,h,c\}$. This is the distinctive feature of a \textit{quantum tricycle} \cite{1310.0683v1} (see Fig.~\ref{fig1}). In absence of heat leaks or internal friction, a quantum tricycle exchanges quanta with all three baths at a single stationary rate $\mathscr{I}$, i.e. $\dot{\mathcal{Q}}_\alpha=-\omega_\alpha\mathscr{I}_\alpha$, with $\mathcal{I}_h=-\mathcal{I}_c=-\mathcal{I}_w\equiv\mathcal{I}$ (in what follows $\hbar=k_B=1$). Thence, the fulfilment of the first law in Eq.~\eqref{first_law} demands to tune the filters in resonance so that $\omega_w=\omega_h-\omega_c$.

Such `ideal' devices have two complementary modes of operation compatible with Eq.~\eqref{second_law}: The absorption/compression refrigerator $\{\dot{\mathcal{Q}}_w>0,\dot{\mathcal{Q}}_h<0,\dot{\mathcal{Q}}_c>0\}$ and the heat transformer/heat engine $\{\dot{\mathcal{Q}}_w<0,\dot{\mathcal{Q}}_h>0,\dot{\mathcal{Q}}_c<0\}$ \cite{andresen1976tricycle,Gordon2000}. Let us consider for instance a compression refrigerator $(T_w\rightarrow\infty)$ at fixed $\omega_h$, for which the inequality \eqref{second_law} may be rewritten as $\omega_c\leq\omega_{c,\,\text{rev}}\equiv \omega_h T_c/T_h$. As $\omega_c\rightarrow\omega_{c,\,\text{rev}}$, the contact ports simultaneously reach local thermal equilibrium with their respective heat reservoirs, and the COP is maximized $(\varepsilon\rightarrow\varepsilon_C)$ \cite{PhysRev.156.343}. In general, however, the effective temperatures $T_\alpha'$ defined from the stationary state of the contacts, do not coincide with the corresponding equilibrium values $T_\alpha$, and the COP is strictly smaller than $\varepsilon_C$.

The irreversibility hindering the cooling performance of ideal quantum tricycles might be thus understood as if only arising from imperfect thermal contact with the heat baths. It is in this sense that we refer to them as `endoreversible'. Alternatively, ideal energy conversion systems may be tagged `strongly coupled' \cite{TF9656101897,esposito2009universality}, referring to the fact that their energy fluxes remain at all times proportional to each other. This is a necessary prerequisite for any device to achieve maximum efficiency, although at vanishing energy-conversion rates \cite{TF9656101897}.

\section{Optimal COP for large temperatures}\label{sec3}

Next, we shall tune the frequency filters of a generic endoreversible power-driven tricycle in the refrigerator configuration, so as to maximize its cooling power in search for the optimal COP. 

From Eq.~\eqref{second_law} it follows that the entropy production can be written as $\dot{S}=x_h\mathscr{I}_h + x_c\mathscr{I}_c$, where the fluxes are $\mathscr{I}_h=-\mathscr{I}_c\equiv\mathscr{I}$, and their conjugate thermodynamic forces are given by $x_\alpha\equiv\omega_\alpha/T_\alpha$. Note that refrigeration is achieved whenever $x_c<x_h$, according to Eq.~\eqref{second_law}. Even though we shall concentrate on the dependence of the flux on the thermodynamic forces, it will generally be a function of other independent dimensionless combinations of parameters, describing the system-bath interactions and the spectrum of thermal fluctuations of the heat reservoirs. 

The cold heat current writes as $\vert\dot{\mathcal{Q}}_c\vert= T_c x_c \mathscr{I}(x_h,x_c)$ and its local maximization with respect to $x_c$ at fixed $x_h$, follows from
\begin{equation}
x_{c,*}\frac{\partial\mathscr{I}}{\partial x_c}(x_h,x_{c,*})+\mathscr{I}(x_h,x_{c,*})=0.
\label{optimization}
\end{equation}

Little more can be said without disclosing the full Hamiltonian of the tricycle, except if one restricts to a certain regime of parameters. Here, we shall take the high-temperature limit ($x_\alpha\rightarrow 0$), where e.g.~symmetric quantum heat engines are known to operate at the Yvon-Novikov-Curzon-Ahlborn efficiency \cite{geva1996quantum,1406.6788v1}, and where different models of absorption refrigerators achieve their maximal performance \cite{Correa2013,Correa2014}.

We shall thus approximate $\mathscr{I}(x_h,x_c)$ around $x_\alpha=0$, retaining only the first non-zero term in its Taylor expansion
\begin{equation}
\mathscr{I}(x_h,x_c)=\sum_i \left(\frac{\partial\mathscr{I}}{\partial x_i}\right)_{\vec{0}} x_i + \sum_{ij} \left(\frac{\partial^2\mathscr{I}}{\partial x_i\partial x_j}\right)_{\vec{0}} x_i x_j+\cdots,
\label{expansion}
\end{equation}
and express the optimal `cold force' as $x_{c,*}\simeq C x_h$, to first order in $x_h$. The coefficient $C$ may be obtained by substituting the approximated current of Eq.~\eqref{expansion} into Eq.~\eqref{optimization}, and will thus depend explicitly on the partial derivatives of the stationary heat current evaluated in $x_\alpha=0$. Noting that the COP of an endoreversible refrigerator writes as
\begin{equation}
\varepsilon=\frac{\omega_c}{\omega_h-\omega_c}=\left(\frac{\varepsilon_C+1}{\varepsilon_C}\frac{x_h}{x_c}-1\right)^{-1},
\label{COP_forces}
\end{equation}
the optimal performance, normalized by $\varepsilon_C$, is finally
\begin{equation}
\frac{\varepsilon_*}{\varepsilon_C}=\frac{C}{(1-C)\,\varepsilon_C+1}.
\label{optimal_COP}
\end{equation}
Here, $C$ must be positive and upper-bounded by $1$, so that $0\leq\varepsilon_*\leq\varepsilon_C$. In general, it will be a function of parameters such as the dissipation rate ($\gamma$), ohmicity ($s$), high frequency cutoff ($\Omega_c$), dimensionality of the baths ($d$) or their equilibrium temperatures (through $\varepsilon_C$). Thus, and unlike Eq.~\eqref{curzon_ahlborn}, $\varepsilon_*/\varepsilon_C$ converges to $C(\varepsilon_C=0,\gamma,s,\Omega_c,d,\cdots)$ as $\varepsilon_C\rightarrow 0$, rather than to a \textit{universal} constant value.

The above discussion can be compared to the one done in Ref.~\cite{esposito2009universality} for a generic heat engine in the linear regime: There, the first order term in the expansion of the optimal force $x_{c,*}$ (in that case, around $x_c-x_h\rightarrow 0$) contributed to $\eta_*/\eta_C$ with a universal constant value of $1/2$, while the second order term added a correction, explicitly involving the first and second order partial derivatives of the heat current. In contrast, as we have just seen, the optimal cooling performance is already non-universal to the lowest order in $x_h$.

In order to intuitively understand this fundamental difference between engines and refrigerators, we remark that the useful effect in a heat engine is sought at the interface of the working substance with an infinite-temperature heat reservoir, implying that the corresponding contact transitions will be saturated regardless of the details of the system-bath interaction. On the contrary, in a refrigerator, the useful effect takes place in the interface with a bath at some finite temperature. Therefore, it is not so surprising that the spectral properties of the environmental fluctuations play a relevant role in establishing the optimal cooling performance. 

Indeed, the situation resembles that of the maximization of the cooling power of endoreversible (`classical') refrigerators, for which the optimal performance is generally set by the heat conductivities, and depends critically on the specific heat transfer law assumed \cite{0022-3727_23_2_002}.

Finally, let us comment on the optimal COP in the complementary limit of $\varepsilon_C\rightarrow\infty$, that is, in the linear regime. Close to equilibrium, we may assume a linear relation between fluxes $\mathcal{I}_\alpha$ and forces $x_\alpha$, such that $\mathcal{I}_h=L_{11}x_h + L_{12}x_c$ and $\mathcal{I}_c=L_{21}x_h + L_{22}x_c$. The Onsager coefficients $L_{ij}$ satisfy $L_{11}\geq 0$, $L_{22}\geq 0$, $L_{12}=L_{21}$ and $q^2\equiv L_{12}^2/L_{11}L_{22}$. Here, the parameter $-1\leq q\leq 1$ is stands for the tightness of the coupling between input and output fluxes \cite{TF9656101897}, where $q^2\rightarrow 1$ implies `endoreversiblity'. We can maximize again $\vert\dot{\mathcal{Q}}_c\vert=T_c x_c\mathcal{I}_c$ in $x_c$ for fixed $x_h$, obtaining $x_{c,*}=-L_{21}x_h / 2 L_{22}$. This yields an optimal COP of
\begin{equation}
\varepsilon_*=\frac{q^2 \varepsilon_C}{(4-3q^2)\varepsilon_C+(4-2q^2)},
\end{equation}
which converges to $\varepsilon_*=q^2/(4-3q^2)$ as $\varepsilon_C\rightarrow\infty$. Hence, the ratio $\varepsilon_*/\varepsilon_C$ simply vanishes close to equilibrium, \textit{regardless} of the magnitude of $x_\alpha$ and the details system-bath coupling.

\section{Example: Unstructured bosonic baths}\label{sec4}

In order to get a closed expression for $C(\varepsilon_C,\gamma,s,\Omega_c,d,\cdots)$, specific instances have to be considered. Here we focus on a simple and paradigmatic endoreversible device, such as a three-level maser \cite{PhysRevLett.2.262} subject to a weak periodic driving, in contact with unstructured bosonic baths (i.e.~characterized by a flat spectral density) in $d_\alpha$ dimensions. Its Hamiltonian writes as
\begin{equation}
H=\omega_c\ket{2}\bra{2}+\omega_h\ket{3}\bra{3}+\lambda\left(e^{i\omega_w t}\ket{2}\bra{3}+e^{-i\omega_w t}\ket{3}\bra{2}\right),
\label{Hamiltonian}
\end{equation}
where $\lambda$ is the intensity of the driving at the power input transition $\ket{2}\leftrightarrow\ket{3}$. The remaining ones ($\ket{1}\leftrightarrow\ket{3}$ and $\ket{1}\leftrightarrow\ket{2}$), are linearly connected with the `hot' and `cold' heat reservoirs, through terms of the form $\sigma_\alpha\otimes\mathcal{B}_\alpha$, where
\begin{subequations}
\begin{align}
\mathcal{B}_\alpha &\equiv\sum\nolimits_\mu g_{\alpha\mu} \left(b_{\alpha\mu}+b_{\alpha\mu}^\dagger\right) \\
\sigma_h &\equiv\ket{1}\bra{3}+\ket{3}\bra{1} \\
\sigma_c &\equiv\ket{1}\bra{2}+\ket{2}\bra{1}.
\label{sys-bath}\end{align}
\end{subequations}
The constants $g_{\alpha\mu}\propto(\gamma_\alpha\omega_{\mu})^{1/2}$ indicate the intensity of the coupling between the mode $\omega_\mu$ of bath $\alpha$ and the corresponding contact transition of the working substance, and $\gamma_\alpha$ stands for the dissipation strength \cite{Correa2013}.

We shall assume very weak dissipation (i.e. $\gamma_\alpha\ll T_\alpha$) and parameters well into the quantum optical regime, so as to consistently derive a quantum master equation like $\dot{\varrho}=\sum_\alpha\sum_\omega\sum_{q\in\mathbb{Z}}\,\mathcal{L}^\alpha_{\omega,q}\,\varrho$, with dissipators $\mathcal{L}^\alpha_{\omega,q}$ of the Lindblad-Gorini-Kossakowski-Sudarshan type \cite{lindblad1976generators,gorini1976completely}. Their explicit form is given in Appendix \ref{appendix}.

The non-equilibrium \textit{limit cycle} state $\varrho_\infty$ may be found from $\sum_\alpha\sum_\omega\sum_{q\in\mathbb{Z}}\,\mathcal{L}^\alpha_{\omega,q}\,\varrho_\infty=0$, while the corresponding (time-averaged) heat currents are \cite{1205.4552v1}
\begin{equation}
\dot{\mathcal{Q}}_\alpha=-T_\alpha\sum_\omega\sum_{q\in\mathbb{Z}}\text{tr}\{\mathcal{L}^\alpha_{\omega,q}\,\varrho_\infty\ln{\tilde{\varrho}^\alpha_{\omega,q}}\}.
\label{heat_currents}
\end{equation}
The states $\tilde{\varrho}^\alpha_{\omega,q}$ are the unique local stationary states of each dissipator, i.e. $\mathcal{L}^\alpha_{\omega,q}\,\tilde{\varrho}^\alpha_{\omega,q}=0$.

In general, a power-driven three-level maser does not realize a tricycle as it features closed performance characteristics, which is a clear indicator of irreversibility \cite{1310.0683v1}. We shall take, however, the limit of weak driving, i.e. $\lambda\rightarrow 0$, in which the time averaged limit flux $\mathscr{I}$ reads
\begin{equation}
\mathscr{I}\simeq\frac{\Gamma_{\omega_h}\Gamma_{-\omega_c}-\Gamma_{-\omega_h}\Gamma_{\omega_c}}{\Gamma_{\omega_h}+\Gamma_{\omega_c}+2\big(\Gamma_{-\omega_h}+\Gamma_{-\omega_c}\big)}.
\label{limit_current}
\end{equation}
The excitation and relaxation rates $\Gamma_{\pm\omega_\alpha}$ are given by $\Gamma_{\omega_\alpha}\equiv\gamma_\alpha\omega^{d_\alpha}[N(\omega_\alpha)+1]$ and $\Gamma_{-\omega_\alpha}=e^{-\omega_\alpha/T_\alpha}\Gamma_{\omega_\alpha}$, with $N(\omega_\alpha)\equiv(e^{\omega_\alpha/T_\alpha}-1)^{-1}$. Here, $d_\alpha$ stands for the physical dimensionality of bath $\alpha$ \cite{breuer2002theory}.

Taking now the high-temperature limit would result in $\Gamma_{\omega_\alpha}\simeq\gamma_\alpha T_\alpha \omega_\alpha^{d_\alpha-1}$ and $\Gamma_{-\omega_\alpha}\simeq\gamma_\alpha T_\alpha \omega_\alpha^{d_\alpha-1}(1-x_\alpha)$, so that
\begin{equation}
\mathscr{I}\simeq\frac{\gamma_h\gamma_c}{3}T_c \,\omega_c^{d_c-1}\frac{\omega_h/T_h-\omega_c/T_c}{\gamma_h+ \omega_c^{d_c-1}\omega_h^{1-d_h} \gamma_c T_c/T_h}.
\label{limit_current_approx}\end{equation}
We shall discard the second term in the denominator of Eq.~\eqref{limit_current_approx} by assuming that the coupling to the entropy sink is much stronger than the interaction with the cold bath (i.e. $\gamma_c\ll\gamma_h$). Setting up a comparatively efficient heat rejection mechanism is indeed very important for the maximization of the stationary flux in a refrigerator, which justifies this assumption as a first step towards optimality. Nonetheless, noting that $T_c/T_h=\varepsilon_C/(\varepsilon_C+1)$, we see that this would be justified anyway, as long as $\varepsilon_C\rightarrow 0$. The stationary flux may be thus written as
\begin{equation}
\mathscr{I}\simeq\mathscr{I}_0 (x_c^{d_c-1}x_h-x_c^{d_c}),
\label{limit_current_approx_2}\end{equation}
with $\mathscr{I}_0=\gamma_c T_c^{d_c}/3$. From here, it follows that $C \equiv d_c/(d_c+1)$, i.e.~$x_{c,*}=d_c/(d_c+1)\,\omega_h$, which once substituted in Eq.~(\ref{optimal_COP}) yields the following simple performance benchmark,
\begin{equation}
\frac{\varepsilon_*}{\varepsilon_C}=\frac{d_c}{d_c+1+\varepsilon_C}.
\label{perf_bound}\end{equation}

\begin{figure}
	\includegraphics[scale=0.42]{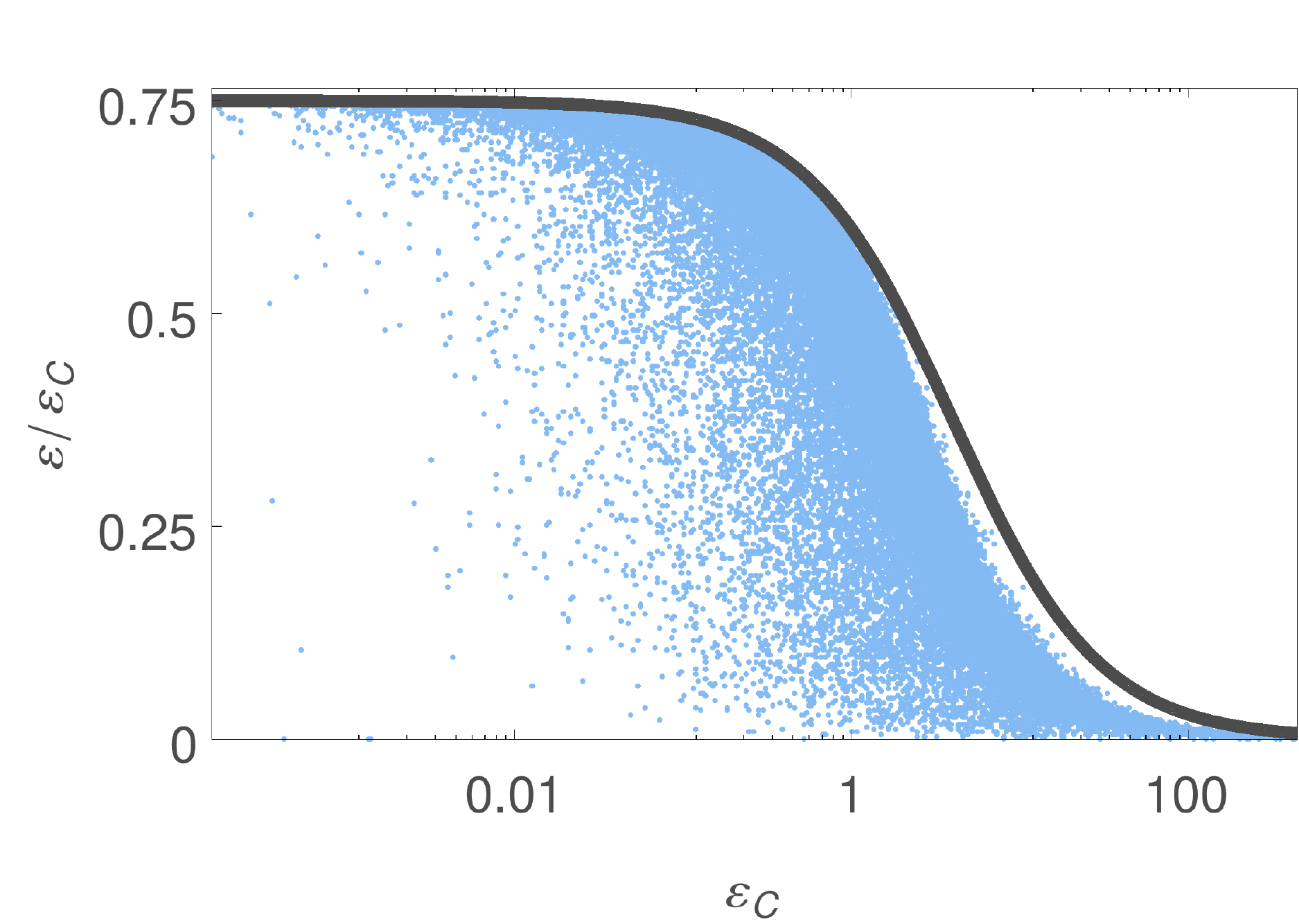}
\caption{(Blue dots) Optimal normalized COP versus $\varepsilon_C$ for about $2 \times 10^{5}$ $n$--stage endoreversible absorption refrigerators \cite{PhysRevE.89.042128} with $n\in\{1,\cdots,10\}$, and coupled to unstructured three-dimensional bosonic baths. All three temperatures $T_\alpha$, dissipation rates $\gamma_\alpha$ and hot frequencies $\omega_h$ were picked at random, and the COP was optimized in $\omega_c$ so as to maximize the cooling power $\dot{\mathcal{Q}}_c$ in each case. Eq.~\eqref{perf_bound} is plotted in solid gray.}
\label{fig4}
\end{figure}

In Fig.~\ref{fig4}, the optimal normalized COP of a large number of single- and multi-stage endoreversible absorption refrigerators \cite{PhysRevE.89.042128} is compared with Eq.~\eqref{perf_bound}, considering unstructured bosonic baths with dimensionality $d_\alpha=3$. We observe a remarkable agreement, especially at low $\varepsilon_C$.
Notice, however, that Eq.~\eqref{perf_bound} was obtained for a specific model of compression refrigerator \footnote{Alternatively, we can consider an absorption three-level maser, driven by heat from a third reservoir at $T_w$ \cite{PhysRevE.64.056130} and then, take the limit $T_w\rightarrow\infty$. Eq.~\eqref{limit_current_approx} would be thus exactly reproduced.} and under the assumption of asymmetric dissipation: There is, in principle, no reason, why it should remain \textit{tight} nor an upper bound to the performance of other endoreversible models. Therefore, it should be thought-of just as a reasonable approximation to their generic behaviour. On a second thought, however, the excellent agreement observed may not be so surprising, provided that the optimal performance is set by the dissipative scheme alone. Solving for the limit cycle of a (weakly-driven) compression tree-level maser in different types of environment would be thus enough to come up with generally valid benchmarks for any endoreversible refrigerator in each case. This is one of the take-home messages of the present paper.

The optimal performance of single and multi-stage quantum absorption refrigerators is indeed known to be limited by $\varepsilon_*/\varepsilon_C\leq d/(d+1)$ when attached to unstructured baths in $d$ dimensions, with a saturation occurring precisely in the limit of large temperatures \cite{Correa2014,PhysRevE.89.042128}.  Eq.~\eqref{perf_bound} can be thus regarded for these models as a stronger bound which sharpens the one given in Ref.~\cite{PhysRevE.89.042128} for any finite $\varepsilon_C$. Remarkably, also another model of non-ideal refrigerator, with the same dissipative scheme, has been shown to have an optimal performance below $\varepsilon_*/\varepsilon_C=d/(d+1)$ \cite{Correa2013}. Note, however that Eq.~(\ref{perf_bound}) should not be expected to hold quantitatively (and not even as a qualitative indicator of optimality) when moving away from endoreversibility.

\section{Conclusions}\label{sec5}

To summarize and conclude, we have shown from first principles how the COP at maximum cooling power of endoreversible quantum tricycles is \textit{not} universal in the high-temperature limit, but fundamentally constrained by the details of their interaction with the external heat reservoirs. For quantum refrigerators coupled to unstructured bosonic baths, we obtained a compact expression for their optimal performance, only dependent on the Carnot COP \textit{and} the dimensionality of the baths.

Our results highlight the importance of reservoir engineering \cite{PhysRevLett.77.4728} in the design of quantum thermal devices: While squeezed-thermal and other types of engineered non-equilibrium environments are known to be capable of enhancing both the performance and power of heat engines and quantum refrigerators \cite{PhysRevE.86.051105,1303.6558v1,PhysRevLett.112.030602,Correa2014}, the exploration of more exotic and highly tunable reservoirs, such as cold atomic gases \cite{0295-5075-101-6-60005}, might bring about new possibilities for the physical realization of super-efficient thermodynamic cycles, especially interesting for practical applications to quantum technologies.

\section*{Acknowledgements}
The authors are grateful to R. Uzdin and A. Levy for fruitful discussions and constructive criticism. This project was funded by COST Action MP1209, the Spanish MICINN (Grant No. FIS2010-19998), by the University of Nottingham through an Early Career Research and Knowledge Transfer Award and an EPSRC Research Development Fund Grant (PP-0313/36), and by the Brazilian funding agency CAPES (Pesquisador Visitante Especial-Grant No. 108/2012).

\appendix

\section{Master equation for a periodically-driven three-level maser}\label{appendix}

In what follows, we shall derive a quantum master equation for a three-level maser weakly coupled to two unstructured bosonic reservoirs in $d$ dimensions, and driven by a periodic perturbation. As already stated in the main text, the full Hamiltonian of system and baths (excluding their mutual interactions) is given by
\begin{multline}
H_0(t)=H(t)+H_B=\omega_c\ket{2}\bra{2}+\omega_h\ket{3}\bra{3}+\\\lambda\left(e^{i\omega_w t}\ket{2}\bra{3}+
e^{-i\omega_w t}\ket{3}\bra{2}\right)+\sum_{\alpha=\{h,c\}}\sum_\lambda\omega_\lambda b_{\alpha\lambda}^{\dagger}b_{\alpha\lambda},
\label{H_0}
\end{multline}
while the system-bath coupling writes as
\begin{equation}
H_I=\sum_{\alpha=\{h,c\}}\sigma_\alpha\otimes\left(\sum_\lambda g_{\alpha\lambda} (b_{\alpha\lambda}+b^{\dagger}_{\alpha\lambda})\right).
\label{H_I}
\end{equation}
Recall that the `thermal contact' operators $\sigma_\alpha$ were just $\sigma_h \equiv\ket{1}\bra{3}+\ket{3}\bra{1}$ and $\sigma_c \equiv\ket{1}\bra{2}+\ket{2}\bra{1}$. The standard \textit{recipe} to derive a Lindbland-Gorini-Kossakovsky-Sudarshan quantum master equation \cite{breuer2002theory} demands to express the two $\sigma_\alpha$ in the interaction picture with respect to $H_0(t)$, and then, to suitably decompose them. 
\newline

In the present case, the unitary evolution operator associated with $H_0(t)$ is formally given by the time-ordered exponential $U_0(t)=\mathcal{T}\exp\left\lbrace-i \int_0^t ds~H_0(s)\right\rbrace$, and may be written as $U_0(t) = U_1(t) U_2(t)\otimes e^{-i H_B t}$, where
\begin{subequations}
\begin{align}
U_1(t)&\equiv\exp{\left\lbrace-i t \left(\omega_c\ket{2}\bra{2}+\omega_h\ket{3}\bra{3}\right)\right\rbrace} \label{U_1}\\ 
U_2(t)&\equiv\exp{\left\lbrace-i t \lambda \left(\ket{2}\bra{3}+\omega_h\ket{3}\bra{2}\right)\right\rbrace}.
\label{unitary_decomposition}
\end{align}
\end{subequations}
This may be easily checked by noticing that $\frac{d}{dt}\left\lbrace U_1(t)U_2(t)\right\rbrace=-i H(t) U_1(t)U_2(t)$. 
\newline

A time-independent (or time-averaged) Hamiltonian $\bar{H}$ can be defined, that generates the same unitary dynamics as $U(t)$ [i.e. $e^{-i\bar{H}t}\equiv U_1(t)U_2(t)$]. For our three-level maser, this would be
\begin{equation}
\bar{H}=\omega_c\left(\ket{2}\bra{2}+\ket{3}\bra{3}\right)+\lambda\left(\ket{2}\bra{3}+\ket{3}\bra{2}\right),
\label{time_averaged_Hamiltonian}
\end{equation}
with eigenvalues $\epsilon=\{0,\omega_c-\lambda,\omega_c+\lambda\}$. Its corresponding set of positive Bohr quasi-frequencies ($\epsilon_j-\epsilon_i>0$) is thus $\bar{\omega}=\{0,2\lambda,\omega_c \pm \lambda\}$, where we have assumed without loss of generality that $\omega_c>\lambda$. 
\newline

In general, we would have to resort now to Floquet theory \cite{1205.4552v1,PhysRevE.87.012120} in order to decompose the interaction picture thermal contact operators as 
\begin{equation}
U(t)^\dagger\sigma_\alpha U(t)=\sum_{\bar{\omega}}\sum_{q\in\mathbb{Z}} A_{\bar{\omega},q}^\alpha e^{-i(\bar{\omega}+q\omega_w)t}. 
\label{decomposition}
\end{equation}
Fortunately for us, this may be done by mere inspection of the left-hand side of Eq.~\eqref{decomposition}, resulting in
\begin{subequations}
\begin{align}
A^h_{\omega_c+\lambda,1}&=\frac{1}{2}\left(\ket{1}\bra{3}+\ket{1}\bra{2}\right) \\ A^h_{\omega_c-\lambda,1}&=\frac{1}{2}\left(\ket{1}\bra{3}-\ket{1}\bra{2}\right) \\
A^c_{\omega_c+\lambda,0}&=\frac{1}{2}\left(\ket{1}\bra{2}+\ket{1}\bra{3}\right) \\ A^c_{\omega_c-\lambda,0}&=\frac{1}{2}\left(\ket{1}\bra{2}-\ket{1}\bra{3}\right) 
\end{align}
\vspace{-0.46cm}
\begin{equation}
A^\alpha_{-\bar{\omega},-q}={A^\alpha_{\bar{\omega},q}}^{\dagger}.
\end{equation}
\label{jump_operators}
\end{subequations}
There are, therefore, two open decay channels for each thermal contact, corresponding to frequencies $\omega_\alpha\pm \lambda$ ($\omega_w=\omega_h-\omega_c$).
\newline

Provided with the decomposition of Eq.~\eqref{jump_operators}, we can now successively apply the Born, Markov and rotating-wave (or secular) approximations on the effective equation of motion of the reduced density operator of the system in the interaction picture $\varrho(t)$ \cite{1205.4552v1,breuer2002theory}. We thus arrive to a quantum master equation in the standard form:
\begin{multline}
\frac{d\varrho}{dt}=\sum_\alpha\sum_{\bar{\omega}}\sum_{q\in\mathbb{Z}} \mathcal{L}_{\bar{\omega},q}^\alpha[\varrho]\\
\equiv\sum_\alpha\sum_{\bar{\omega}}\sum_{q\in\mathbb{Z}}\left[ \Gamma_{\bar{\omega},q}^{\alpha}\left(A_{\bar{\omega},q}^{\alpha}\varrho{A_{\bar{\omega},q}^{\alpha}}^\dagger-\frac{1}{2}\{{A_{\bar{\omega},q}^{\alpha}}^\dagger A_{\bar{\omega},q}^{\alpha},\varrho\}_+\right)+\right.\\
\left.\Gamma_{-\bar{\omega},-q}^{\alpha}\left({A_{\bar{\omega},q}^{\alpha}}^\dagger\varrho A_{\bar{\omega},q}^{\alpha}-\frac{1}{2}\{A_{\bar{\omega},q}^{\alpha} {A_{\bar{\omega},q}^{\alpha}}^\dagger,\varrho\}_+\right)\right].
\label{Lindblad_equation}
\end{multline}
The assumption of factorized initial conditions between system and environmental degrees of freedom is implicit in the above, as is thermal equilibrium for the hot and cold heat reservoirs. Also note that the Lamb-shift term has been neglected in Eq.~\eqref{Lindblad_equation}. 
\newline

The relaxation rates $\Gamma_{\omega}^\alpha=2\text{Re}\left\lbrace\int_{0}^\infty ds~e^{i\omega s} \left\langle B_\alpha(t) B_\alpha(t-s) \right\rangle\right\rbrace$ are determined by the power spectrum of the environmental fluctuations, and satisfy the Kubo-Martin-Schwinger condition \cite{kubo1957statistical,PhysRev.115.1342} $\Gamma_{-\omega}^\alpha=e^{-\omega/T_\alpha}\Gamma_\omega^\alpha$. Here, $\left\langle\cdots\right\rangle$ stands for equilibrium averaging. As already advanced in the main text, for our choice of the system-baths coupling scheme, i.e. bosonic baths with \textit{constant} spectral density $J_\alpha(\omega)\sim\gamma_\alpha$, the relaxation rates are explicitly given by $\Gamma_{\omega}^\alpha=\gamma_\alpha\omega^{d_\alpha}[N_\alpha(\omega)+1]$, with $N_\alpha(\omega)\equiv(e^{\omega/T_\alpha}-1)^{-1}$. Physically, this is compatible with weak coupling to the quantized electromagnetic field in thermal equilibrium, inside a $d_\alpha$--dimensional box \cite{breuer2002theory}.
\newline

Equipped with Eqs.~\eqref{jump_operators} and \eqref{Lindblad_equation}, we are now in the position of finding the \textit{limit cycle} state $\varrho_\infty$, which is defined as
\begin{equation}
\sum_\alpha\sum_{\bar{\omega}}\sum_{q\in\mathbb{Z}} \mathcal{L}_{\bar{\omega},q}^\alpha[\varrho_\infty]=0.
\label{limit_cycle}\end{equation}

The dissipators $\mathcal{L}_{\bar{\omega},q}^\alpha$ have local steady states (i.e. $\mathcal{L}^\alpha_{\bar{\omega},q}[\tilde{\varrho}_{\bar{\omega},q}^\alpha]=0$) of the form $\tilde{\varrho}^\alpha_{\bar{\omega},q}=Z^{-1}\exp\left\lbrace-\frac{\bar{\omega}+q\omega_w}{\bar{\omega}}\bar{H}\right\rbrace$ \cite{1205.4552v1}. Given their standard Lindblad form, each $\mathcal{L}^\alpha_{\bar{\omega},q}$ individually generates a fully \textit{contractive} reduced dynamics towards $\tilde{\varrho}^\alpha_{\bar{\omega},q}$, which is reflected in the monotonic decrease of the distance, as measured by the relative entropy, from any locally evolved state $\frac{d}{dt}\varrho(t)=\mathcal{L}^\alpha_{\bar{\omega},q}[\varrho(t)]$ to $\tilde{\varrho}^\alpha_{\bar{\omega},q}$ [i.e. $\frac{d}{dt}S(\varrho(t)\vert\vert\tilde{\varrho}^\alpha_{\omega,q})\leq 0$] \cite{spohn1978entropy,breuer2002theory}. Such contractivity property applied to the actual steady state of the full Eq.~\eqref{Lindblad_equation} eventually leads to the following inequality \cite{1205.4552v1}
\begin{equation}
\sum_\alpha\frac{1}{T_\alpha}\left(-T_\alpha\sum_{\bar{\omega}}\sum_{q\in\mathbb{Z}}\text{Tr}~\mathcal{L}^\alpha_{\bar{\omega},q}[\varrho_\infty]\log\tilde{\varrho}_{\bar{\omega},q}^\alpha\right)\leq 0,
\label{second_law1}\end{equation}
or equivalently
\begin{equation}
\sum_\alpha\frac{1}{T_\alpha}\left(\sum_{\bar{\omega}}\sum_{q\in\mathbb{Z}}\frac{\bar{\omega}+q\omega_w}{\bar{\omega}}~\text{Tr}~\bar{H}\mathcal{L}^\alpha_{\bar{\omega},q}[\varrho_\infty]\right)\leq 0.
\label{second_law2}\end{equation}
This can be understood as a statement of the second law of thermodynamics upon defining the limit cycle heat currents as $\dot{\mathcal{Q}}_\alpha\equiv\sum_{\bar{\omega}}\sum_{q\in\mathbb{Z}}\frac{\bar{\omega}+q\omega_w}{\bar{\omega}}~\text{Tr}\{\bar{H}\mathcal{L}^\alpha_{\bar{\omega},q}[\varrho_\infty]\}$ \cite{1205.4552v1}. 

As it is probably useful for the interested reader, we now detail the specific form of the hot and cold dissipators. These are given by
\begin{eqnarray}\label{eq:dissipators}
{\cal L}_{\omega_\alpha+\lambda}^\alpha[\varrho]&=&
\frac{\Gamma_{\omega_\alpha+\lambda}}{4}{\cal D}_{++}[\varrho]\,+\,
\frac{\Gamma_{-\omega_\alpha-\lambda}}{4}{\cal D}_{--}[\varrho]\,,
\nonumber\\
{\cal L}_{\omega_\alpha-\lambda}^\alpha[\varrho]&=&
\frac{\Gamma_{\omega_\alpha-\lambda}}{4}{\cal D}_{+-}[\varrho]\,+\,
\frac{\Gamma_{-\omega_\alpha+\lambda}}{4}{\cal D}_{-+}[\varrho]\,,
\end{eqnarray}
To simplify the notation, we have introduced the superoperators ${\cal D}$, which act on $\varrho$ as
\begin{eqnarray}
{\cal D}_{++}[\varrho]&=&
(|1\rangle\langle 2|+|1\rangle\langle 3|)\varrho
(|2\rangle\langle 1|+|3\rangle\langle 1|)
\nonumber\\
&-&\frac{1}{2}(|2\rangle\langle 2|+|2\rangle\langle 3|
+|3\rangle\langle 2|+|3\rangle\langle 3|)\varrho
\nonumber\\
&-&\frac{1}{2}\varrho(|2\rangle\langle 2|+|2\rangle\langle 3|
+|3\rangle\langle 2|+|3\rangle\langle 3|)\,,
\nonumber\\
{\cal D}_{--}[\varrho]&=&(|2\rangle\langle 1|+|3\rangle\langle 1|)\varrho
(|1\rangle\langle 2|+|1\rangle\langle 3|)
\nonumber\\
&-&|1\rangle\langle 1|\varrho-\varrho|1\rangle\langle 1|\,,
\nonumber\\
{\cal D}_{+-}[\varrho]&=&(|1\rangle\langle 2|-|1\rangle\langle 3|)\varrho
(|2\rangle\langle 1|-|3\rangle\langle 1|)
\nonumber\\
&-&\frac{1}{2}(|2\rangle\langle 2|-|2\rangle\langle 3|
-|3\rangle\langle 2|+|3\rangle\langle 3|)\varrho
\nonumber\\
&-&\frac{1}{2}\varrho(|2\rangle\langle 2|-|2\rangle\langle 3|
-|3\rangle\langle 2|+|3\rangle\langle 3|)\,,
\nonumber\\
{\cal D}_{-+}[\varrho]&=&(|2\rangle\langle 1|-|3\rangle\langle 1|)\varrho
(|1\rangle\langle 2|-|1\rangle\langle 3|)
\nonumber\\
&-&|1\rangle\langle 1|\varrho-\varrho|1\rangle\langle 1|\,.
\end{eqnarray}

The populations of the limit cycle state $\varrho_\infty$ (expressed in vector form as $\mathbf{n}=\{n_1,n_2,n_3\}$) may be found by combining the relation ${\bold M}^{D3L}\cdot{\bf n}=\mathbf{0}$ with the normalisation condition $\sum_i\,n_i=1$. The coefficient matrix ${\bold M}^{D3L}$ is given by
\begin{equation}
{\bold M}^{D3L}=
\left(\begin{array}{ccc}
-2G^+_{-\omega}+2\frac{G^-_{\omega}G^-_{-\omega}}{G^+_{\omega}} 
& 
G^+_{\omega}-\frac{G^-_{\omega}G^-_{\omega}}{G^+_{\omega}} 
& 
G^+_{\omega}-\frac{G^-_{\omega}G^-_{\omega}}{G^+_{\omega}}
\\
G^+_{-\omega}-\frac{G^-_{\omega}G^-_{-\omega}}{G^+_{\omega}} 
&
-G^+_{\omega}+\frac{G^-_{\omega}G^-_{\omega}}{2G^+_{\omega}}  
&
\frac{G^-_{\omega}G^-_{\omega}}{2G^+_{\omega}}
\\
G^+_{-\omega}-\frac{G^-_{\omega}G^-_{-\omega}}{G^+_{\omega}}
&
\frac{G^-_{\omega}G^-_{\omega}}{2G^+_{\omega}}
& 
-G^+_{\omega}+\frac{G^-_{\omega}G^-_{\omega}}{2G^+_{\omega}}
\end{array}\right)\,,
\label{matrix}\end{equation}
where the constants $G^\pm_{\omega}$ are defined as
\begin{equation}
\begin{array}{cc} 
G^+_\omega\,\equiv\,\frac{\Gamma^S_{\omega+\lambda}+\Gamma^S_{\omega-\lambda}}{4}\,,
&
G^+_{-\omega}\,\equiv\,\frac{\Gamma^S_{-\omega-\lambda}+\Gamma^S_{-\omega+\lambda}}{4}\,,
\\
G^-_\omega\,\equiv\,\frac{\Gamma^S_{\omega+\lambda}-\Gamma^S_{\omega-\lambda}}{4}\,,
&
G^-_{-\omega}\,\equiv\,\frac{\Gamma^S_{-\omega-\lambda}-\Gamma^S_{-\omega+\lambda}}{4}\,,
\end{array},
\end{equation}
and $\Gamma^S_{\pm\omega\pm\lambda}\,\equiv\,
\Gamma^\alpha_{\pm\omega_h\pm\lambda}\,+\,\Gamma^\alpha_{\pm\omega_c\pm\lambda}$.
\newline

Finally, we also give the explicit form of the cycle-averaged stationary heat flows. In particular, $\dot{\mathcal{Q}}_h$ is given by 
\begin{eqnarray}
&&\dot{\cal Q}_h\,=\,\dot{\cal Q}_{\omega_h+\lambda}\,+\,\dot{\cal Q}_{\omega_h-\lambda}\,,
\nonumber\\
&&\dot{\cal Q}_{\omega_h+\lambda}\,=\,\frac{\omega_h+\lambda}{\omega_c+\lambda}
\,{\rm Tr}\{\bar{H}{\cal L}_{\omega_h+\lambda}[\tilde{\rho}]\}=\,
\nonumber\\
&&(\omega_h+\lambda)\left[
\frac{\Gamma_{-\omega_h-\lambda}}{2}n_1\,-\,
\frac{\Gamma_{\omega_h+\lambda}}{4}(n_2+n_3+n_c)\right]\,,
\nonumber\\
&&\dot{\cal Q}_{\omega_h-\lambda}\,=\,\frac{\omega_h-\lambda}{\omega_c-\lambda}
\,{\rm Tr}\{\bar{H}{\cal L}_{\omega_h-\lambda}[\tilde{\rho}]\}\,=\,
\nonumber\\
&&(\omega_h-\lambda)\left[
\frac{\Gamma_{-\omega_h+\lambda}}{2}n_1\,-\,
\frac{\Gamma_{\omega_h-\lambda}}{4}(n_2+n_3-n_c)\right]\,,
\nonumber
\label{hot_current}\end{eqnarray}
and $\dot{\mathcal{Q}}_c$ writes as
\begin{eqnarray}
&&\dot{\cal Q}_c\,=\,\dot{\cal Q}_{\omega_c+\lambda}\,+\,\dot{\cal Q}_{\omega_c-\lambda}\,,
\nonumber\\
&&\dot{\cal Q}_{\omega_c+\lambda}\,=\,
{\rm Tr}\{\bar{H}{\cal L}_{\omega_c+\lambda}[\tilde{\rho}]\}
\nonumber\\
&&=(\omega_c+\lambda)\left[
\frac{\Gamma_{-\omega_c-\lambda}}{2}n_1\,-\,
\frac{\Gamma_{\omega_c+\lambda}}{4}(n_2+n_3+n_c)\right]\,,
\nonumber\\
&&\dot{\cal Q}_{\omega_c-\lambda}\,=\,
{\rm Tr}\{\bar{H}{\cal L}_{\omega_c-\lambda}[\tilde{\rho}]\}
\nonumber\\
&&=(\omega_c-\lambda)\left[
\frac{\Gamma_{-\omega_c+\lambda}}{2}n_1\,-\,
\frac{\Gamma_{\omega_c-\lambda}}{4}(n_2+n_3-n_c)\right]\,.
\nonumber
\label{cold_current}\end{eqnarray}
In these expressions, the constant $n_c$ is defined as
\begin{equation}
n_c\,\equiv\,(n_{23}+n_{32})\,=\,2\frac{G^-_{-\omega}}{G^+_{\omega}}\,n_1\,-\,
\frac{G^-_{\omega}}{G^+_{\omega}}(n_2+n_3)\,,
\end{equation}
where $n_{ij}=\langle i|\varrho_\infty|j\rangle$ stands for steady-state coherences.
\newline

Getting the steady-state populations from Eq.~\eqref{matrix} and using the xpressions for the heat currents above, allows to check the validity of Eq.~\eqref{limit_current} in the limit of $\lambda\rightarrow 0$.

\bibliographystyle{apsrev}

\begin{thebibliography}{43}
\expandafter\ifx\csname natexlab\endcsname\relax\def\natexlab#1{#1}\fi
\expandafter\ifx\csname bibnamefont\endcsname\relax
  \def\bibnamefont#1{#1}\fi
\expandafter\ifx\csname bibfnamefont\endcsname\relax
  \def\bibfnamefont#1{#1}\fi
\expandafter\ifx\csname citenamefont\endcsname\relax
  \def\citenamefont#1{#1}\fi
\expandafter\ifx\csname url\endcsname\relax
  \def\url#1{\texttt{#1}}\fi
\expandafter\ifx\csname urlprefix\endcsname\relax\def\urlprefix{URL }\fi
\providecommand{\bibinfo}[2]{#2}
\providecommand{\eprint}[2][]{\url{#2}}

\bibitem[{\citenamefont{Andresen}(2011)}]{andresen2011current}
\bibinfo{author}{\bibfnamefont{B.}~\bibnamefont{Andresen}},
  \bibinfo{journal}{Angewandte Chemie International Edition}
  \textbf{\bibinfo{volume}{50}}, \bibinfo{pages}{2690} (\bibinfo{year}{2011}).

\bibitem[{\citenamefont{Hoffmann et~al.}(1997)\citenamefont{Hoffmann, Burzler,
  and Schubert}}]{hoffmann1997endoreversible}
\bibinfo{author}{\bibfnamefont{K.~H.} \bibnamefont{Hoffmann}},
  \bibinfo{author}{\bibfnamefont{J.~M.} \bibnamefont{Burzler}},
  \bibnamefont{and} \bibinfo{author}{\bibfnamefont{S.}~\bibnamefont{Schubert}},
  \bibinfo{journal}{J. Non-Equilib. Thermodyn} \textbf{\bibinfo{volume}{22}},
  \bibinfo{pages}{311} (\bibinfo{year}{1997}).

\bibitem[{\citenamefont{Novikov}(1957)}]{novikov1957efficiency}
\bibinfo{author}{\bibfnamefont{I.}~\bibnamefont{Novikov}},
  \bibinfo{journal}{Atomic Energy} \textbf{\bibinfo{volume}{3}},
  \bibinfo{pages}{1269} (\bibinfo{year}{1957}).

\bibitem[{\citenamefont{Carnot}(1890)}]{carnot1890fire}
\bibinfo{author}{\bibfnamefont{S.}~\bibnamefont{Carnot}},
  \emph{\bibinfo{title}{Reflections on the Motive Power of Heat and on Machines
  Fitted to Develop That Power}} (\bibinfo{publisher}{J. Wiley \& Sons (New
  York)}, \bibinfo{year}{1890}).

\bibitem[{\citenamefont{Yvon}(1955)}]{Yvon1955reactor}
\bibinfo{author}{\bibfnamefont{J.}~\bibnamefont{Yvon}}, in
  \emph{\bibinfo{booktitle}{Proceedings of the International Conference on
  Peaceful Uses of Atomic Energy (United Nations, Geneva)}}
  (\bibinfo{year}{1955}), p. \bibinfo{pages}{387}.

\bibitem[{\citenamefont{Curzon and Ahlborn}(1975)}]{curzon1975efficiency}
\bibinfo{author}{\bibfnamefont{F.}~\bibnamefont{Curzon}} \bibnamefont{and}
  \bibinfo{author}{\bibfnamefont{B.}~\bibnamefont{Ahlborn}},
  \bibinfo{journal}{Am. J. Phys.} \textbf{\bibinfo{volume}{43}},
  \bibinfo{pages}{22} (\bibinfo{year}{1975}).

\bibitem[{\citenamefont{Vaudrey et~al.}()\citenamefont{Vaudrey, Lanzetta, and
  Feidt}}]{1406.5853v2}
\bibinfo{author}{\bibfnamefont{A.}~\bibnamefont{Vaudrey}},
  \bibinfo{author}{\bibfnamefont{F.}~\bibnamefont{Lanzetta}}, \bibnamefont{and}
  \bibinfo{author}{\bibfnamefont{M.}~\bibnamefont{Feidt}},
  \bibinfo{note}{e-print arXiv:1406.5853}.

\bibitem[{\citenamefont{Van~den Broeck}(2005)}]{PhysRevLett.95.190602}
\bibinfo{author}{\bibfnamefont{C.}~\bibnamefont{Van~den Broeck}},
  \bibinfo{journal}{Phys. Rev. Lett.} \textbf{\bibinfo{volume}{95}},
  \bibinfo{pages}{190602} (\bibinfo{year}{2005}).

\bibitem[{\citenamefont{Esposito et~al.}(2009)\citenamefont{Esposito,
  Lindenberg, and Van~den Broeck}}]{esposito2009universality}
\bibinfo{author}{\bibfnamefont{M.}~\bibnamefont{Esposito}},
  \bibinfo{author}{\bibfnamefont{K.}~\bibnamefont{Lindenberg}},
  \bibnamefont{and} \bibinfo{author}{\bibfnamefont{C.}~\bibnamefont{Van~den
  Broeck}}, \bibinfo{journal}{Phys. Rev. Lett.} \textbf{\bibinfo{volume}{102}},
  \bibinfo{pages}{130602} (\bibinfo{year}{2009}).

\bibitem[{\citenamefont{Esposito
  et~al.}(2010{\natexlab{a}})\citenamefont{Esposito, Kawai, Lindenberg, and
  Van~den Broeck}}]{PhysRevLett.105.150603}
\bibinfo{author}{\bibfnamefont{M.}~\bibnamefont{Esposito}},
  \bibinfo{author}{\bibfnamefont{R.}~\bibnamefont{Kawai}},
  \bibinfo{author}{\bibfnamefont{K.}~\bibnamefont{Lindenberg}},
  \bibnamefont{and} \bibinfo{author}{\bibfnamefont{C.}~\bibnamefont{Van~den
  Broeck}}, \bibinfo{journal}{Phys. Rev. Lett.} \textbf{\bibinfo{volume}{105}},
  \bibinfo{pages}{150603} (\bibinfo{year}{2010}{\natexlab{a}}).

\bibitem[{\citenamefont{Esposito
  et~al.}(2010{\natexlab{b}})\citenamefont{Esposito, Kawai, Lindenberg, and
  Van~den Broeck}}]{esposito2010universalityCA}
\bibinfo{author}{\bibfnamefont{M.}~\bibnamefont{Esposito}},
  \bibinfo{author}{\bibfnamefont{R.}~\bibnamefont{Kawai}},
  \bibinfo{author}{\bibfnamefont{K.}~\bibnamefont{Lindenberg}},
  \bibnamefont{and} \bibinfo{author}{\bibfnamefont{C.}~\bibnamefont{Van~den
  Broeck}}, \bibinfo{journal}{Phys. Rev. E} \textbf{\bibinfo{volume}{81}},
  \bibinfo{pages}{041106} (\bibinfo{year}{2010}{\natexlab{b}}).

\bibitem[{\citenamefont{Geva and Kosloff}(1992)}]{geva1991spin}
\bibinfo{author}{\bibfnamefont{E.}~\bibnamefont{Geva}} \bibnamefont{and}
  \bibinfo{author}{\bibfnamefont{R.}~\bibnamefont{Kosloff}},
  \bibinfo{journal}{The Journal of chemical physics}
  \textbf{\bibinfo{volume}{96}}, \bibinfo{pages}{3054} (\bibinfo{year}{1992}).

\bibitem[{\citenamefont{Yan and Chen}(1990)}]{0022-3727_23_2_002}
\bibinfo{author}{\bibfnamefont{Z.}~\bibnamefont{Yan}} \bibnamefont{and}
  \bibinfo{author}{\bibfnamefont{J.}~\bibnamefont{Chen}},
  \bibinfo{journal}{Journal of Physics D: Applied Physics}
  \textbf{\bibinfo{volume}{23}}, \bibinfo{pages}{136} (\bibinfo{year}{1990}).

\bibitem[{\citenamefont{de~Tom\'as et~al.}(2012)\citenamefont{de~Tom\'as,
  Hern\'andez, and Roco}}]{PhysRevE.85.010104}
\bibinfo{author}{\bibfnamefont{C.}~\bibnamefont{de~Tom\'as}},
  \bibinfo{author}{\bibfnamefont{A.~C.} \bibnamefont{Hern\'andez}},
  \bibnamefont{and} \bibinfo{author}{\bibfnamefont{J.~M.~M.}
  \bibnamefont{Roco}}, \bibinfo{journal}{Phys. Rev. E}
  \textbf{\bibinfo{volume}{85}}, \bibinfo{pages}{010104}
  (\bibinfo{year}{2012}).

\bibitem[{\citenamefont{Velasco et~al.}(1997)\citenamefont{Velasco, Roco,
  Medina, and Hern\'andez}}]{PhysRevLett.78.3241}
\bibinfo{author}{\bibfnamefont{S.}~\bibnamefont{Velasco}},
  \bibinfo{author}{\bibfnamefont{J.~M.~M.} \bibnamefont{Roco}},
  \bibinfo{author}{\bibfnamefont{A.}~\bibnamefont{Medina}}, \bibnamefont{and}
  \bibinfo{author}{\bibfnamefont{A.~C.} \bibnamefont{Hern\'andez}},
  \bibinfo{journal}{Phys. Rev. Lett.} \textbf{\bibinfo{volume}{78}},
  \bibinfo{pages}{3241} (\bibinfo{year}{1997}).

\bibitem[{\citenamefont{Jim\'enez~de Cisneros
  et~al.}(2006)\citenamefont{Jim\'enez~de Cisneros, Arias-Hern\'andez, and
  Hern\'andez}}]{PhysRevE.73.057103}
\bibinfo{author}{\bibfnamefont{B.}~\bibnamefont{Jim\'enez~de Cisneros}},
  \bibinfo{author}{\bibfnamefont{L.~A.} \bibnamefont{Arias-Hern\'andez}},
  \bibnamefont{and} \bibinfo{author}{\bibfnamefont{A.~C.}
  \bibnamefont{Hern\'andez}}, \bibinfo{journal}{Phys. Rev. E}
  \textbf{\bibinfo{volume}{73}}, \bibinfo{pages}{057103}
  (\bibinfo{year}{2006}).

\bibitem[{\citenamefont{Chen}(1994)}]{chen1994new}
\bibinfo{author}{\bibfnamefont{J.}~\bibnamefont{Chen}},
  \bibinfo{journal}{Journal of Physics A: Mathematical and General}
  \textbf{\bibinfo{volume}{27}}, \bibinfo{pages}{6395} (\bibinfo{year}{1994}).

\bibitem[{\citenamefont{Apertet et~al.}(2013)\citenamefont{Apertet, Ouerdane,
  Michot, Goupil, and Lecoeur}}]{0295-5075-103-4-40001}
\bibinfo{author}{\bibfnamefont{Y.}~\bibnamefont{Apertet}},
  \bibinfo{author}{\bibfnamefont{H.}~\bibnamefont{Ouerdane}},
  \bibinfo{author}{\bibfnamefont{A.}~\bibnamefont{Michot}},
  \bibinfo{author}{\bibfnamefont{C.}~\bibnamefont{Goupil}}, \bibnamefont{and}
  \bibinfo{author}{\bibfnamefont{P.}~\bibnamefont{Lecoeur}},
  \bibinfo{journal}{EPL (Europhysics Letters)} \textbf{\bibinfo{volume}{103}},
  \bibinfo{pages}{40001} (\bibinfo{year}{2013}).

\bibitem[{\citenamefont{Kosloff and Levy}(2014)}]{1310.0683v1}
\bibinfo{author}{\bibfnamefont{R.}~\bibnamefont{Kosloff}} \bibnamefont{and}
  \bibinfo{author}{\bibfnamefont{A.}~\bibnamefont{Levy}},
  \bibinfo{journal}{Anual Rev. Phys. Chem.} \textbf{\bibinfo{volume}{65}},
  \bibinfo{pages}{365} (\bibinfo{year}{2014}).

\bibitem[{\citenamefont{Scovil and Schulz-DuBois}(1959)}]{PhysRevLett.2.262}
\bibinfo{author}{\bibfnamefont{H.~E.~D.} \bibnamefont{Scovil}}
  \bibnamefont{and} \bibinfo{author}{\bibfnamefont{E.~O.}
  \bibnamefont{Schulz-DuBois}}, \bibinfo{journal}{Phys. Rev. Lett.}
  \textbf{\bibinfo{volume}{2}}, \bibinfo{pages}{262} (\bibinfo{year}{1959}).

\bibitem[{\citenamefont{Palao et~al.}(2001)\citenamefont{Palao, Kosloff, and
  Gordon}}]{PhysRevE.64.056130}
\bibinfo{author}{\bibfnamefont{J.~P.} \bibnamefont{Palao}},
  \bibinfo{author}{\bibfnamefont{R.}~\bibnamefont{Kosloff}}, \bibnamefont{and}
  \bibinfo{author}{\bibfnamefont{J.~M.} \bibnamefont{Gordon}},
  \bibinfo{journal}{Phys. Rev. E} \textbf{\bibinfo{volume}{64}},
  \bibinfo{pages}{056130} (\bibinfo{year}{2001}).

\bibitem[{\citenamefont{Andresen et~al.}(1977)\citenamefont{Andresen, Salamon,
  and Berry}}]{andresen1976tricycle}
\bibinfo{author}{\bibfnamefont{B.}~\bibnamefont{Andresen}},
  \bibinfo{author}{\bibfnamefont{P.}~\bibnamefont{Salamon}}, \bibnamefont{and}
  \bibinfo{author}{\bibfnamefont{R.~S.} \bibnamefont{Berry}},
  \bibinfo{journal}{The Journal of Chemical Physics}
  \textbf{\bibinfo{volume}{66}}, \bibinfo{pages}{1571} (\bibinfo{year}{1977}).

\bibitem[{\citenamefont{Gordon and Ng}(2000)}]{Gordon2000}
\bibinfo{author}{\bibfnamefont{J.~M.} \bibnamefont{Gordon}} \bibnamefont{and}
  \bibinfo{author}{\bibfnamefont{K.~C.} \bibnamefont{Ng}},
  \emph{\bibinfo{title}{Cool thermodynamics}} (\bibinfo{publisher}{Cambridge
  international science publishing Cambridge}, \bibinfo{year}{2000}).

\bibitem[{\citenamefont{Geusic et~al.}(1967)\citenamefont{Geusic,
  Schulz-DuBios, and Scovil}}]{PhysRev.156.343}
\bibinfo{author}{\bibfnamefont{J.~E.} \bibnamefont{Geusic}},
  \bibinfo{author}{\bibfnamefont{E.~O.} \bibnamefont{Schulz-DuBios}},
  \bibnamefont{and} \bibinfo{author}{\bibfnamefont{H.~E.~D.}
  \bibnamefont{Scovil}}, \bibinfo{journal}{Phys. Rev.}
  \textbf{\bibinfo{volume}{156}}, \bibinfo{pages}{343} (\bibinfo{year}{1967}).

\bibitem[{\citenamefont{Kedem and Caplan}(1965)}]{TF9656101897}
\bibinfo{author}{\bibfnamefont{O.}~\bibnamefont{Kedem}} \bibnamefont{and}
  \bibinfo{author}{\bibfnamefont{S.~R.} \bibnamefont{Caplan}},
  \bibinfo{journal}{Trans. Faraday Soc.} \textbf{\bibinfo{volume}{61}},
  \bibinfo{pages}{1897} (\bibinfo{year}{1965}).

\bibitem[{\citenamefont{Geva and Kosloff}(1996)}]{geva1996quantum}
\bibinfo{author}{\bibfnamefont{E.}~\bibnamefont{Geva}} \bibnamefont{and}
  \bibinfo{author}{\bibfnamefont{R.}~\bibnamefont{Kosloff}},
  \bibinfo{journal}{J. Chem. Phys.} \textbf{\bibinfo{volume}{104}},
  \bibinfo{pages}{7681} (\bibinfo{year}{1996}).

\bibitem[{\citenamefont{Raam and Kosloff}()}]{1406.6788v1}
\bibinfo{author}{\bibfnamefont{U.}~\bibnamefont{Raam}} \bibnamefont{and}
  \bibinfo{author}{\bibfnamefont{R.}~\bibnamefont{Kosloff}},
  \bibinfo{note}{e-print arXiv:1406.6788}.

\bibitem[{\citenamefont{Correa et~al.}(2013)\citenamefont{Correa, Palao,
  Adesso, and Alonso}}]{Correa2013}
\bibinfo{author}{\bibfnamefont{L.~A.} \bibnamefont{Correa}},
  \bibinfo{author}{\bibfnamefont{J.~P.} \bibnamefont{Palao}},
  \bibinfo{author}{\bibfnamefont{G.}~\bibnamefont{Adesso}}, \bibnamefont{and}
  \bibinfo{author}{\bibfnamefont{D.}~\bibnamefont{Alonso}},
  \bibinfo{journal}{Phys. Rev. E} \textbf{\bibinfo{volume}{87}},
  \bibinfo{pages}{042131} (\bibinfo{year}{2013}).

\bibitem[{\citenamefont{Correa et~al.}(2014)\citenamefont{Correa, Palao,
  Alonso, and Adesso}}]{Correa2014}
\bibinfo{author}{\bibfnamefont{L.~A.} \bibnamefont{Correa}},
  \bibinfo{author}{\bibfnamefont{J.~P.} \bibnamefont{Palao}},
  \bibinfo{author}{\bibfnamefont{D.}~\bibnamefont{Alonso}}, \bibnamefont{and}
  \bibinfo{author}{\bibfnamefont{G.}~\bibnamefont{Adesso}},
  \bibinfo{journal}{Sci. Rep.} \textbf{\bibinfo{volume}{4}}
  (\bibinfo{year}{2014}).

\bibitem[{\citenamefont{Lindblad}(1976)}]{lindblad1976generators}
\bibinfo{author}{\bibfnamefont{G.}~\bibnamefont{Lindblad}},
  \bibinfo{journal}{Comm. Math. Phys.} \textbf{\bibinfo{volume}{48}},
  \bibinfo{pages}{119} (\bibinfo{year}{1976}).

\bibitem[{\citenamefont{Gorini et~al.}(1976)\citenamefont{Gorini, Kossakowski,
  and Sudarshan}}]{gorini1976completely}
\bibinfo{author}{\bibfnamefont{V.}~\bibnamefont{Gorini}},
  \bibinfo{author}{\bibfnamefont{A.}~\bibnamefont{Kossakowski}},
  \bibnamefont{and}
  \bibinfo{author}{\bibfnamefont{E.}~\bibnamefont{Sudarshan}},
  \bibinfo{journal}{J. Math. Phys.} \textbf{\bibinfo{volume}{17}},
  \bibinfo{pages}{821} (\bibinfo{year}{1976}).

\bibitem[{\citenamefont{Robert~Alicki}(2012)}]{1205.4552v1}
\bibinfo{author}{\bibfnamefont{G.~K.} \bibnamefont{Robert~Alicki},
  \bibfnamefont{David Gelbwaser-Klimovsky}} (\bibinfo{year}{2012}).

\bibitem[{\citenamefont{Breuer and Petruccione}(2002)}]{breuer2002theory}
\bibinfo{author}{\bibfnamefont{H.}~\bibnamefont{Breuer}} \bibnamefont{and}
  \bibinfo{author}{\bibfnamefont{F.}~\bibnamefont{Petruccione}},
  \emph{\bibinfo{title}{The Theory of Open Quantum Systems}}
  (\bibinfo{publisher}{Oxford University Press, USA}, \bibinfo{year}{2002}).

\bibitem[{\citenamefont{Correa}(2014)}]{PhysRevE.89.042128}
\bibinfo{author}{\bibfnamefont{L.~A.} \bibnamefont{Correa}},
  \bibinfo{journal}{Phys. Rev. E} \textbf{\bibinfo{volume}{89}},
  \bibinfo{pages}{042128} (\bibinfo{year}{2014}).

\bibitem[{\citenamefont{Poyatos et~al.}(1996)\citenamefont{Poyatos, Cirac, and
  Zoller}}]{PhysRevLett.77.4728}
\bibinfo{author}{\bibfnamefont{J.~F.} \bibnamefont{Poyatos}},
  \bibinfo{author}{\bibfnamefont{J.~I.} \bibnamefont{Cirac}}, \bibnamefont{and}
  \bibinfo{author}{\bibfnamefont{P.}~\bibnamefont{Zoller}},
  \bibinfo{journal}{Phys. Rev. Lett.} \textbf{\bibinfo{volume}{77}},
  \bibinfo{pages}{4728} (\bibinfo{year}{1996}).

\bibitem[{\citenamefont{Huang et~al.}(2012)\citenamefont{Huang, Wang, and
  Yi}}]{PhysRevE.86.051105}
\bibinfo{author}{\bibfnamefont{X.~L.} \bibnamefont{Huang}},
  \bibinfo{author}{\bibfnamefont{T.}~\bibnamefont{Wang}}, \bibnamefont{and}
  \bibinfo{author}{\bibfnamefont{X.~X.} \bibnamefont{Yi}},
  \bibinfo{journal}{Phys. Rev. E} \textbf{\bibinfo{volume}{86}},
  \bibinfo{pages}{051105} (\bibinfo{year}{2012}).

\bibitem[{\citenamefont{Abah and Lutz}(2014)}]{1303.6558v1}
\bibinfo{author}{\bibfnamefont{O.}~\bibnamefont{Abah}} \bibnamefont{and}
  \bibinfo{author}{\bibfnamefont{E.}~\bibnamefont{Lutz}}, \bibinfo{journal}{EPL
  (Europhysics Letters)} \textbf{\bibinfo{volume}{106}}, \bibinfo{pages}{20001}
  (\bibinfo{year}{2014}).

\bibitem[{\citenamefont{Ro\ss{}nagel et~al.}(2014)\citenamefont{Ro\ss{}nagel,
  Abah, Schmidt-Kaler, Singer, and Lutz}}]{PhysRevLett.112.030602}
\bibinfo{author}{\bibfnamefont{J.}~\bibnamefont{Ro\ss{}nagel}},
  \bibinfo{author}{\bibfnamefont{O.}~\bibnamefont{Abah}},
  \bibinfo{author}{\bibfnamefont{F.}~\bibnamefont{Schmidt-Kaler}},
  \bibinfo{author}{\bibfnamefont{K.}~\bibnamefont{Singer}}, \bibnamefont{and}
  \bibinfo{author}{\bibfnamefont{E.}~\bibnamefont{Lutz}},
  \bibinfo{journal}{Phys. Rev. Lett.} \textbf{\bibinfo{volume}{112}},
  \bibinfo{pages}{030602} (\bibinfo{year}{2014}).

\bibitem[{\citenamefont{McEndoo et~al.}(2013)\citenamefont{McEndoo, Haikka,
  Chiara, Palma, and Maniscalco}}]{0295-5075-101-6-60005}
\bibinfo{author}{\bibfnamefont{S.}~\bibnamefont{McEndoo}},
  \bibinfo{author}{\bibfnamefont{P.}~\bibnamefont{Haikka}},
  \bibinfo{author}{\bibfnamefont{G.~D.} \bibnamefont{Chiara}},
  \bibinfo{author}{\bibfnamefont{G.~M.} \bibnamefont{Palma}}, \bibnamefont{and}
  \bibinfo{author}{\bibfnamefont{S.}~\bibnamefont{Maniscalco}},
  \bibinfo{journal}{EPL (Europhysics Letters)} \textbf{\bibinfo{volume}{101}},
  \bibinfo{pages}{60005} (\bibinfo{year}{2013}).

\bibitem[{\citenamefont{Szczygielski et~al.}(2013)\citenamefont{Szczygielski,
  Gelbwaser-Klimovsky, and Alicki}}]{PhysRevE.87.012120}
\bibinfo{author}{\bibfnamefont{K.}~\bibnamefont{Szczygielski}},
  \bibinfo{author}{\bibfnamefont{D.}~\bibnamefont{Gelbwaser-Klimovsky}},
  \bibnamefont{and} \bibinfo{author}{\bibfnamefont{R.}~\bibnamefont{Alicki}},
  \bibinfo{journal}{Phys. Rev. E} \textbf{\bibinfo{volume}{87}},
  \bibinfo{pages}{012120} (\bibinfo{year}{2013}).

\bibitem[{\citenamefont{Kubo}(1957)}]{kubo1957statistical}
\bibinfo{author}{\bibfnamefont{R.}~\bibnamefont{Kubo}},
  \bibinfo{journal}{Journal of the Physical Society of Japan}
  \textbf{\bibinfo{volume}{12}}, \bibinfo{pages}{570} (\bibinfo{year}{1957}).

\bibitem[{\citenamefont{Martin and Schwinger}(1959)}]{PhysRev.115.1342}
\bibinfo{author}{\bibfnamefont{P.~C.} \bibnamefont{Martin}} \bibnamefont{and}
  \bibinfo{author}{\bibfnamefont{J.}~\bibnamefont{Schwinger}},
  \bibinfo{journal}{Phys. Rev.} \textbf{\bibinfo{volume}{115}},
  \bibinfo{pages}{1342} (\bibinfo{year}{1959}).

\bibitem[{\citenamefont{Spohn}(1978)}]{spohn1978entropy}
\bibinfo{author}{\bibfnamefont{H.}~\bibnamefont{Spohn}},
  \bibinfo{journal}{Journal of Mathematical Physics}
  \textbf{\bibinfo{volume}{19}}, \bibinfo{pages}{1227} (\bibinfo{year}{1978}).

\end{thebibliography}

\end{document}